\documentclass[twocolumn,pra,showpacs]{revtex4}
\usepackage[dvipdfmx]{graphicx}
\usepackage{amsmath}
\usepackage{amssymb}
\usepackage{amsthm}
\usepackage{bm}
\usepackage{latexsym}

\newcommand\ee[0]{\mathrm{e}}

\newcommand\Tz[0]{$T_\mathrm{Z}$ }

\newcommand\xx[0]{\hat x}
\newcommand\pp[0]{\hat p}

\usepackage{textcomp}

\newcommand\refFig[1]{Fig.~\ref{fig:#1}}
\newcommand\refFigsAnd[3]{Figs.~\ref{fig:#1}#2 and \ref{fig:#1}#3}
\newcommand\refFigs[3]{Figs.~\ref{fig:#1}#2--\ref{fig:#1}#3}
\newcommand\refFigure[1]{Figure~\ref{fig:#1}}

\newcommand\refFigures[3]{Figures~\ref{fig:#1}#2--\ref{fig:#1}#3}

\newcommand\PRL[1]{{\it Phys. Rev. Lett.} {\bf #1},}
\newcommand\PRA[1]{{\it Phys. Rev. A.} {\bf #1},}

\usepackage{ulem}
\usepackage{color} 

\begin{document}

\title{
Demonstration of a fully tuneable entangling gate for continuous-variable one-way quantum computation
}

\author
{Shota Yokoyama$^{1,}$\footnote{yokoyama@alice.t.u-tokyo.ac.jp}, Ryuji Ukai$^{1}$, Seiji C. Armstrong$^{1,2}$, Jun-ichi Yoshikawa$^{1}$,
Peter van Loock$^{3}$, and Akira Furusawa$^{1,}$\footnote{akiraf@ap.t.u-tokyo.ac.jp}}

\affiliation{
$^{1}$Department of Applied Physics, School of Engineering,
The University of Tokyo,\\ 7-3-1 Hongo, Bunkyo-ku, Tokyo 113-8656, Japan\\
$^{2}$Centre for Quantum Computation and Communication Technology, Department of Quantum Science,
Research School of Physics and Engineering, The Australian National University, Canberra ACT 0200 Australia\\
$^{3}$Institute for Physics, University of Mainz, D-55128 Mainz, Germany
}

\begin{abstract}
We introduce a fully tuneable entangling gate for continuous-variable one-way quantum computation. We present a proof-of-principle demonstration by propagating two independent optical inputs through a three-mode linear cluster state and applying the gate in various regimes. The genuine quantum nature of the gate is confirmed by verifying the entanglement strength in the output state.
Our protocol can be readily incorporated into efficient multi-mode interaction operations in the context of large-scale one-way quantum computation, as our tuning process is the generalisation of cluster state shaping.
\end{abstract}
\maketitle
\section{Introduction}
The quantum computer promises an impressive speedup in certain problems such as prime factorisation \cite{Nielsen00}.
Measurement-based quantum computation (MBQC) is one approach for processing quantum information, attractive due to its relative ease of use once a suitable resource state has been prepared.
In MBQC, unitary operations are performed via pre-prepared multi-partite entangled resource states, referred to as cluster states \cite{Briegel01,Raussendorf01,Menicucci06}.
Sufficiently large cluster states are first prepared before being appropriately reshaped for any specific operations.
Arbitrary unitary operations are implemented by the precise selection of measurement bases and outcome-dependent feed-forward operations.

To date there have been several demonstrations of MBQC, predominantly in quantum optics.
Optical experiments performed in a continuous-variable (CV) setting benefit from deterministic state generation as well as deterministic implementations of Gaussian operations. The cluster states that facilitate MBQC can be generated via linear optics \cite{Su2007,Yukawa08,Yokoyama13,Moran13}. Four-mode and six-mode cluster states have been already used to implement arbitrary single-mode Gaussian gates~\cite{Ukai11}, a two-mode Gaussian gate~\cite{Ukai11QND}, and a gate sequence of these two~\cite{Su13}. Reshaping a cluster state \cite{Miwa10} is possible through quantum erasing \cite{Filip03} and wire-shortening \cite{Gu09}, which correspond to erasing and preserving the interaction gains between the nodes of the cluster state, respectively. Recently, large-scale \cite{Moran13} and ultra-large-scale \cite{Yokoyama13} cluster states have been generated by multiplexing in the frequency and time domain, respectively, both based on the same theoretical proposal \cite{Menicucci08,Menicucci11}.

Present techniques for shaping a cluster are inherently inefficient due to the lack of control over the interaction strength. For example the fixed-strength entangling gate demonstrated in Ref.~\cite{Ukai11QND} cannot have its entanglement strength tuned, and therefore it cannot completely make use of the underlying structure of the cluster state \cite{Ukai10}.

In this paper, we present a fully tuneable entangling gate for CV one-way quantum computation and experimentally demonstrate a proof-of-principle implementation. Our tuneable gate can be interpreted as a generalized instance of cluster state reshaping, which we name cluster gain tuning. Our implementation involves propagating two independent quantised optical modes (qumodes) through a three-mode linear cluster state while implementing the gate at various different strengths. The tuneable interaction gain in the resource cluster state is teleported onto the two-mode input state \cite{Bartlett03}, thus appearing at the output and becoming manifest as a certain form of entanglement.

\begin{figure}[b!]
		\centering
		\includegraphics[scale=1.4,clip]{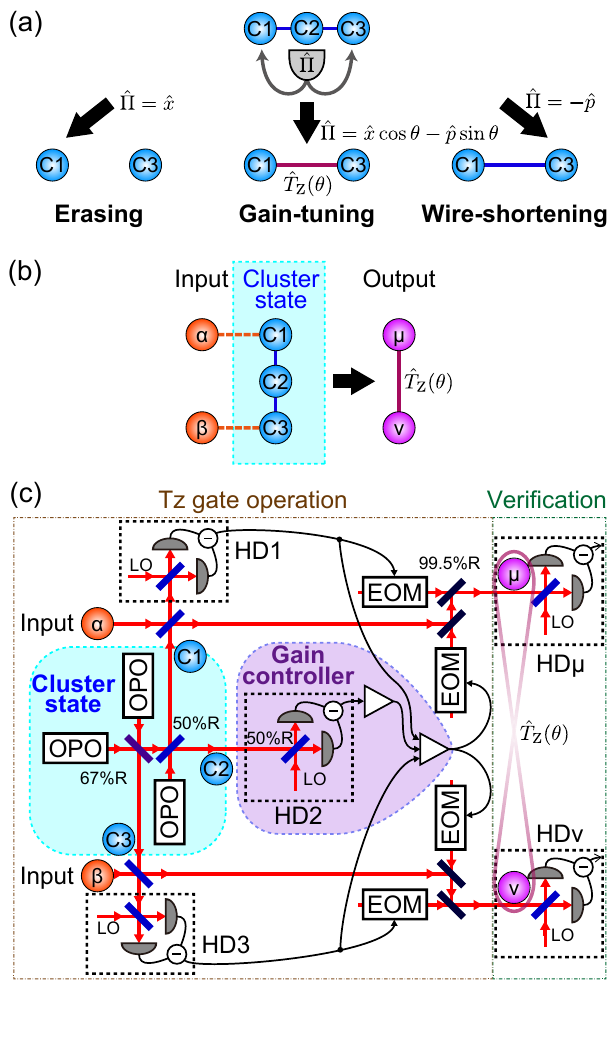}
			\caption{
(Color online)
(a) Cluster state shaping. (top) An initial three-mode linear cluster state. (bottom left) Quantum erasing. (bottom center) Interaction gain tuning. (bottom right) Wire-shortening.
Hemispheric objects next to the nodes (circles) and arrows mean measurements and feedforwards regarding to measurement outcomes, respectively.
 $\hat \Pi$ is the measurement observable.
(b) Abstract illustration of our experiment. Dashed lines represent beam-splitter coupling.
(c) Schematic of our experimental setup.
OPO, optical parametric oscillator;
HD, homodyne detector;
EOM, electro-optical modulator;
$r$\%R, $r$\%-reflectivity beam-splitter;
LO, local oscillator.
}
			\label{fig:ExperimentalSetup}
\end{figure}

\section{protocol for tuneable entangling gates via cluster gain tuning}
Our quantum states are represented by the quadrature operators $(\xx_j, \pp_j )$ of an electric field (annihilation) operator $\hat a_j = \xx_j + i \pp_j$, where the subscript $j$ denotes the $j$-th optical mode.
These quadrature operators play the roles of position and momentum operators of the corresponding harmonic oscillator, and hence they are canonically conjugate variables:
$[\xx_j, \pp_k]=i/2\,\delta_{jk}\ (\hbar =1/2)$,  where $\delta_{jk}$ is the Kronecker delta.
A CV cluster state is defined, in the ideal case, through its zero eigenvalues for certain linear combinations of the canonical operators, so-called nullifiers, 
\begin{align}
\pp_{Cj}-\sum_{k\in N_j} \xx_{Ck} \equiv \hat \delta_{j},
\end{align}
 where $N_j$ refers to the nearest-neighbour nodes of node $j$ in the sense of a general graph \cite{Menicucci11Graph}.
Arbitrary bonds in CV cluster states are generated by applying controlled-phase gates $\hat C_{\mathrm{Z}jk}=\ee^{2i\xx_j \xx_k}$ on pairs of nodes, which are initialised as momentum eigenstates with zero eigenvalues in the limit of infinite squeezing \cite{Menicucci06}.
This can be understood mathematically as the transformation of nullifiers,
\begin{align}
\label{Clusternullifier}
\Big(\sum_{k\in N_j}\hat C_{\mathrm{Z}jk}\Big)\hat p_{Cj}\Big(\sum_{k\in N_j}\hat C_{\mathrm{Z}jk}^\dagger\Big)=\hat \delta_{j}.
\end{align}
The controlled-phase gates will be generalized
to include arbitrary, real gain values, $\hat C_{\mathrm{Z}jk}(g)\equiv\ee^{2ig\xx_j\xx_k}$, leading to certain weighted (real-valued) graph states, with $g=1$ as the special case of unweighted graph states.
More generally, any physical graph state can be described by complex weights and a complex adjacency matrix (including self-loops), 
corresponding to a set of non-Hermitian nullifiers, where the eigenvalue (nullifier) conditions are still
exactly fulfilled even for finite squeezing \cite{Menicucci11Graph}.
However, instead of complex weights in the following,
we describe physical, finitely squeezed cluster states allowing non-zero excess noise in the Hermitian nullifier operators $\hat \delta_{j}$ \cite{Ukai11QND}.

After the preparation of a generic cluster state, the undesired bonds and nodes of the cluster can be erased by means of measurement and feed-forward, applying the quantum eraser \cite{Filip03}.
For example, the three-mode linear cluster state shown in the top of \refFig{ExperimentalSetup}(a), which is  the resource state for our demonstration of the tuneable entangling gate, has bonds ($C1$--$C2$) and ($C2$--$C3$).
By measuring the position operator of node $C2$ ($\xx_{C2}$) and subtracting the measurement outcome from the momentum operators of the nearest-neighbour nodes ($C1$ and $C3$), the bonds are erased and the two modes end up in a separable state [the bottom left of \refFig{ExperimentalSetup}(a)]. 
On the other hand, a node can be deleted while keeping the bond up to local phase rotations, which is called wire-shortening \cite{Gu09}.
By measuring the momentum operator of node $C2$ ($\pp_{C2}$) in the same three-mode linear cluster state and subtracting the measurement outcome from the position operator of a nearest-neighbour node (either $C1$ or $C3$), the resulting two-mode state becomes an Einstein-Podolsky-Rosen (EPR) state or a two-mode cluster state up to local phase rotations [the bottom right of \refFig{ExperimentalSetup}(a)].
The two procedures described above can then be regarded as two extreme cases of the cluster shaping.
Here we generalise these procedures by considering intermediate operations between them, where we can tune the cluster gain between two cluster nodes via the measurement of the center node up to local unitaries.

We now discuss our implementation of cluster gain-tuning on a three-mode linear cluster state.
Here we consider a measurement of the observable $\xx_{C2}\cos\theta-\pp_{C2}\sin\theta$ on cluster node $C2$, where $\theta=0^\circ$ and $90^\circ$ correspond to erasing and wire-shortening, respectively. By subtracting the measurement outcome rescaled by $1/\cos\theta$ from the momentum operators of nearest-neighbour nodes ($C1$ and $C3$), the nullifiers of the resulting state become
\begin{subequations}
\begin{align}
\hat \delta_1'\equiv\pp_{C1}-(\xx_{C1}+\xx_{C3})\tan\theta\\
 \mathrm{and}\quad
\hat \delta_3'\equiv\pp_{C3}-(\xx_{C1}+\xx_{C3})\tan\theta
\end{align}
\end{subequations}
[the bottom center of \refFig{ExperimentalSetup}(a)] (see Appendix B).
In analogy with Eq.\eqref{Clusternullifier}, they correspond to the transformation of nullifiers:
\begin{subequations}
\begin{align}
\label{Tznullifier1}
\hat T_{\mathrm{Z}C1C3}(\theta )\hat p_{C1} \hat T^\dagger_{\mathrm{Z}C1C3}(\theta )&=\hat \delta_1'\\
\label{Tznullifier2}
\hat T_{\mathrm{Z}C1C3}(\theta )\hat p_{C3} \hat T^\dagger_{\mathrm{Z}C1C3}(\theta )&=\hat \delta_3',
\end{align}
\end{subequations}
where the definition of
the unitary operator is
\begin{align}
\hat T_{\mathrm{Z}jk}(\theta)\equiv \ee^{i(\xx_j +\xx_k)^2\tan\theta },
\end{align}
therefore the resulting state corresponds to the application of the gate on two momentum eigenstates with zero eigenvalues.
We name this operation the {\it fully tuneable entangling gate} $T_\mathrm{Z}$, which has the tunable interaction parameter $\tan\theta$.
Since the measurement angle $\theta$ can be set arbitrarily from $-90^\circ$ to $90^\circ$, the \Tz gate can have an arbitrary real value of the interaction parameter $\tan\theta$.
The \Tz gate consists of two quadratic phase gates for individual modes ($\ee^{i\xx_j^2\tan\theta},\ee^{i\xx_k^2\tan\theta} $) \cite{Miwa09} and a controlled-phase gate ($C_{\mathrm{Z}jk}(\tan\theta)=\ee^{2i\xx_j\xx_k\tan\theta}$)
with the arbitrary interaction parameter $\tan \theta$.
The above cluster gain tuning allows for the generation of weighted gain cluster states from larger unweighted cluster states, while additional single-mode operations can be absorbed in the measurements at the latter process in order to perform larger one-way quantum computations.

The tuneable entangling gate is constructed by combining the cluster gain tuning scheme with two input states as shown in \refFig{ExperimentalSetup}(b).
Two input states in modes $\alpha$ and $\beta$ are teleported to modes $C1$ and $C3$ by half Bell measurements and cluster gain tuning, resulting in the \Tz gate operation being teleported onto the input states \cite{Bartlett03}.

In the following we describe the above procedure taking into account the excess noises $\hat \delta_j$ due to finite squeezing.
Each input mode ($\alpha$ or $\beta$) is coupled with a side mode in the cluster state via a balanced beam-splitter (50\%-BS).
Then one output arm of each of the two mixing beam-splitters as well as the centre mode in the cluster state are measured by means of homodyne detection.
The measured observables correspond to
\begin{subequations}
\begin{align}
\hat s_1&\equiv\hat x_{\alpha'}=\tfrac{1}{\sqrt{2}}(\xx_\alpha-\xx_{C1}), \\
\hat s_3&\equiv\hat x_{\beta'}=\tfrac{1}{\sqrt{2}}(\xx_\beta-\xx_{C3}), \\
\mathrm{and} \quad
\hat s_2(\theta)&\equiv \xx_{C2}\cos\theta-\pp_{C2}\sin\theta,
\end{align}
\end{subequations}
where $\theta$ is the measurement angle of the homodyne detection
on the centre mode.
We use primes to mark the modes after each beam-splitter interaction.
The quadratures of the remaining parts are
\begin{subequations}
\begin{align}
\hat x_{C1'}=\dfrac{1}{\sqrt{2}} (\hat x_\alpha +\hat x_{C1}),
\qquad 
\hat p_{C1'}=\dfrac{1}{\sqrt{2}} (\hat p_\alpha +\hat p_{C1}),
\\
\hat x_{C3'}=\dfrac{1}{\sqrt{2}} (\hat x_\beta +\hat x_{C3}),
\qquad 
\hat p_{C3'}=\dfrac{1}{\sqrt{2}} (\hat p_\beta +\hat p_{C3}).
\end{align}
\end{subequations}
Based on the measurement outcomes, we perform the following feed-forward operations onto the rest of the states:
\begin{subequations}
\begin{align}
&\hat X_{C1'}(\hat s_1) \hat Z_{C1'}\big((\hat s_1+\hat s_3)\tan\theta-\tfrac{\hat s_2(\theta)}{\sqrt{2}\cos\theta}\big)\\
\mathrm{and}\quad&\hat X_{C3'}(\hat s_3) \hat Z_{C3'}\big((\hat s_1+\hat s_3)\tan\theta-\tfrac{\hat s_2(\theta)}{\sqrt{2}\cos\theta}\big),
\end{align}
\end{subequations}
where $\hat X_k(s)=\ee ^{-2is\hat p_k}$ and $\hat Z_k(s)=\ee ^{2is\hat x_k}$ are the Weyl-Heisenberg position and momentum displacement operators on the state labeled by $k$, respectively.
The effects of these displacement operators correspond to additions and subtractions for quadratures [see Appendix A]:
\begin{subequations}
\begin{align}
\notag
\hat x_\mu &\equiv \hat x_{C1'}+\hat s_1 \\
&=\sqrt{2}\hat x_{\alpha}\\
\notag
\hat p_\mu &\equiv \hat p_{C1'}+(\hat s_1+\hat s_3)\tan\theta- \tfrac{\hat s_2(\theta)}{\sqrt{2}\cos\theta}\\
&=\tfrac{1}{\sqrt{2}}\Big[\hat p_\alpha +(\hat x_\alpha +\hat x_\beta)\tan\theta +\hat \delta_1+\hat \delta_2\tan\theta\Big]\\
\notag
\hat x_\nu &\equiv \hat x_{C3'}+\hat s_3 \\
&=\sqrt{2}\hat x_{\beta}\\
\notag
\hat p_\nu &\equiv \hat p_{C3'}+(\hat s_1+\hat s_3)\tan\theta- \tfrac{\hat s_2(\theta)}{\sqrt{2}\cos\theta}\\
&=\tfrac{1}{\sqrt{2}}\Big[\hat p_\beta +(\hat x_\alpha +\hat x_\beta)\tan\theta +\hat \delta_3+\hat \delta_2\tan\theta\Big],
\end{align}
\end{subequations}
where we refer to the two output modes as $\mu$ and $\nu$ in order to distinguish them from the input modes denoted by $\alpha$ and $\beta$.
Consequently, the input-output relation in the Heisenberg picture is given by
\begin{align}
\label{eq:inputoutputrelation}
\hat{\bm{\xi}}_{\mu\nu}
&=
\begin{pmatrix}
\bm{S} & \bm{0} \\
\bm{0} & \bm{S}
\end{pmatrix}
\begin{pmatrix}
\bm{I}+\bm{T}(\theta) & \bm{T}(\theta) \\
\bm{T}(\theta) & \bm{I}+\bm{T}(\theta)
\end{pmatrix}
\hat{\bm{\xi}}_{\alpha\beta}
+\hat{\bm{\delta}}
\\
&=
\Big(\hat S_\alpha  \hat S_\beta \hat T_{\mathrm{Z}\alpha\beta}(\theta) \Big)^\dagger
\hat{\bm{\xi}}_{\alpha\beta}
\Big(\hat S_\alpha  \hat S_\beta \hat T_{\mathrm{Z}\alpha\beta}(\theta) \Big)
+\hat{\bm{\delta}},
\end{align}
where
\begin{align}
\bm{S}=\begin{pmatrix}\sqrt{2} & 0 \\0 & 1/\sqrt{2} \end{pmatrix},
\qquad \bm{T}(\theta)=\begin{pmatrix}0 & 0 \\\tan\theta & 0 \end{pmatrix},
\end{align}
$\bm{I}$ is the $2\times 2$ identity matrix,
and
\begin{align}
\label{delta}
\hat{\bm{\delta}}=
\begin{pmatrix}
0 \\
\hat \delta _1 +\hat \delta _2 \tan\theta \\
0 \\
\delta _3 +\hat \delta _2\tan\theta
\end{pmatrix}
\end{align}
 are excess noise terms for imperfect resource squeezing.
There are local squeezing operations
$\hat S_j=\ee^{-i\ln 2 (\hat x_j\hat p_j+\hat p_j \hat x_j )/2}$ in addition to the teleported \Tz gate.
These $-3.0$~dB $p$-squeezing operations are due to the input coupling with a 50\%-BS.
A half teleportation with a beam-splitter coupling corresponds to a squeezing gate \cite{Filip05,Yoshikawa07}.
Note that it can be eliminated by adding an additional coupling node at the edge of cluster states, by which full quantum teleportation with full Bell measurements is performed into the cluster state instead of half teleporation with half Bell measurement \cite{Ukai10}.

In order to verify the entangling capability of the \Tz gate, we now consider the case where both input states are coherent states.
We evaluate the entanglement with the symplectic eigenvalues $\tilde\lambda_-$ of the partially transposed covariance matrix of the output state \cite{Adesso04}.
This corresponds to the logarithmic negativity, which gives $E_N=\max[0,-\ln(4\tilde\lambda_-)]$ for the case of Gaussian states and which is an entanglement measure invariant under local unitary operations \cite{Vidal02}. The covariance matrix is given by
$V\equiv
\tfrac{1}{2}
\bigl\langle \{
\hat{\bm{\xi}} 
,
\hat{\bm{\xi}} 
\}
\bigr\rangle$,
where $\{\hat{\bm{u}}, \hat{\bm{v}}\}\equiv \hat{\bm{u}} \hat{\bm{v}}^T +\bigl(\hat{\bm{v}} \hat{\bm{u}}^T\bigr)^T$ \cite{Menicucci11Graph}.

For our setup the symplectic eigenvalues become
\begin{align}
\notag
\tilde\lambda_-
&=\frac{1}{4}\Big[
1+2t^2+(2+3t^2)\ee^{-2r}\\
&-\sqrt{4t^2(1+t^2+(2+3t^2)\ee^{-2r})+((1+3t^2)\ee^{-2r})^2}\Big]^{1/2},
\end{align}
where $t=\tan\theta$.
It can be calculated by means of the excess noise terms with $\hat \delta_{1}=\sqrt{2}\ee^{-r}\pp_1^{(0)}$, $\hat \delta_{2}=\sqrt{3}\ee^{-r}\pp_2^{(0)}$, and $\hat \delta_{3}=\frac{1}{\sqrt{2}}\ee^{-r}\pp_1^{(0)}+\sqrt{\frac{3}{2}}\ee^{-r}\pp_3^{(0)}$, where $\ee^{-r}\pp_j^{(0)}$ is a squeezed quadrature of the $j$-th resource squeezed-vacuum mode before the beam-splitter network.
Here we assume that all three modes have the same level of squeezing $r$ for simplicity.
The asymmetric case is easily derived in a similar manner.
The ideal (unphysical) cluster state is obtained in the limit $r\to \infty$.

The positivity under partial transposition (PPT criterion) is a necessary (and sufficient in the case of two-mode Gaussian states) measure for the separability of a state \cite{Simon00}. Thus, the output states of our setup are entangled if $\tilde\lambda_-$ is below 1/4.
Furthermore, the closer to zero $\tilde\lambda_-$ is, the stronger is the entanglement in the output states. With respect to our \Tz gate, $\tilde\lambda_-$ becomes smaller as we increase the interaction parameter $\tan\theta$.

\section{Experiment}
The schematic of our experimental setup is shown in \refFig{ExperimentalSetup}(c).
The light source is a continuous-wave Ti:sapphire laser with a wavelength of 860 nm and a power of about 1.7~W. The quantum states to be processed are qumodes at 1~MHz sidebands of the laser beam.
The resource cluster state is prepared by combining three squeezed vacuum states on two beam splitters, each generated by a subthreshold optical parametric oscillator (OPO).
We mainly employ the experimental techniques described in Ref.~\cite{Yukawa08Tele} for the feed-forward of measurement results through classical channels.
Note that the tuneable interaction parameter $t=\tan\theta$ of the \Tz gate is accessed via the relative phase $\theta$ between the signal beam and the reference local oscillator beam at the homodyne-2 detection station (HD-2). The relative phase is precisely controlled via the voltage sent to a piezoelectric transducer (PZT) attached to a mirror. Squeezing levels of the resource squeezed vacuum states are about $-4.5$~dB. The propagation losses from the OPOs to the homodyne detectors are 3\% to 9\%.
The detectors' quantum efficiencies are 99\%, and the interference visibilities are 96\% on average.

In order to evaluate our gate we measure the powers of the quadratures at the homodyne detectors with a spectrum analyzer. The measured frequency is 1~MHz with a resolution bandwidth of 30~kHz, and video bandwidth of 300~Hz. For each quadrature, 101 data points are taken with a sweep time of 0.05~s, while this is repeated 10 times for averaging. Standard errors in these averaged measurements are less than 0.06~dB. In the case of coherent state inputs, we average over even more measurements, leading to standard errors less than 0.01~dB. Note that no corrections are applied for any experimental losses.

\begin{figure}[t!]
		\centering
		\includegraphics[scale=0.65,clip]{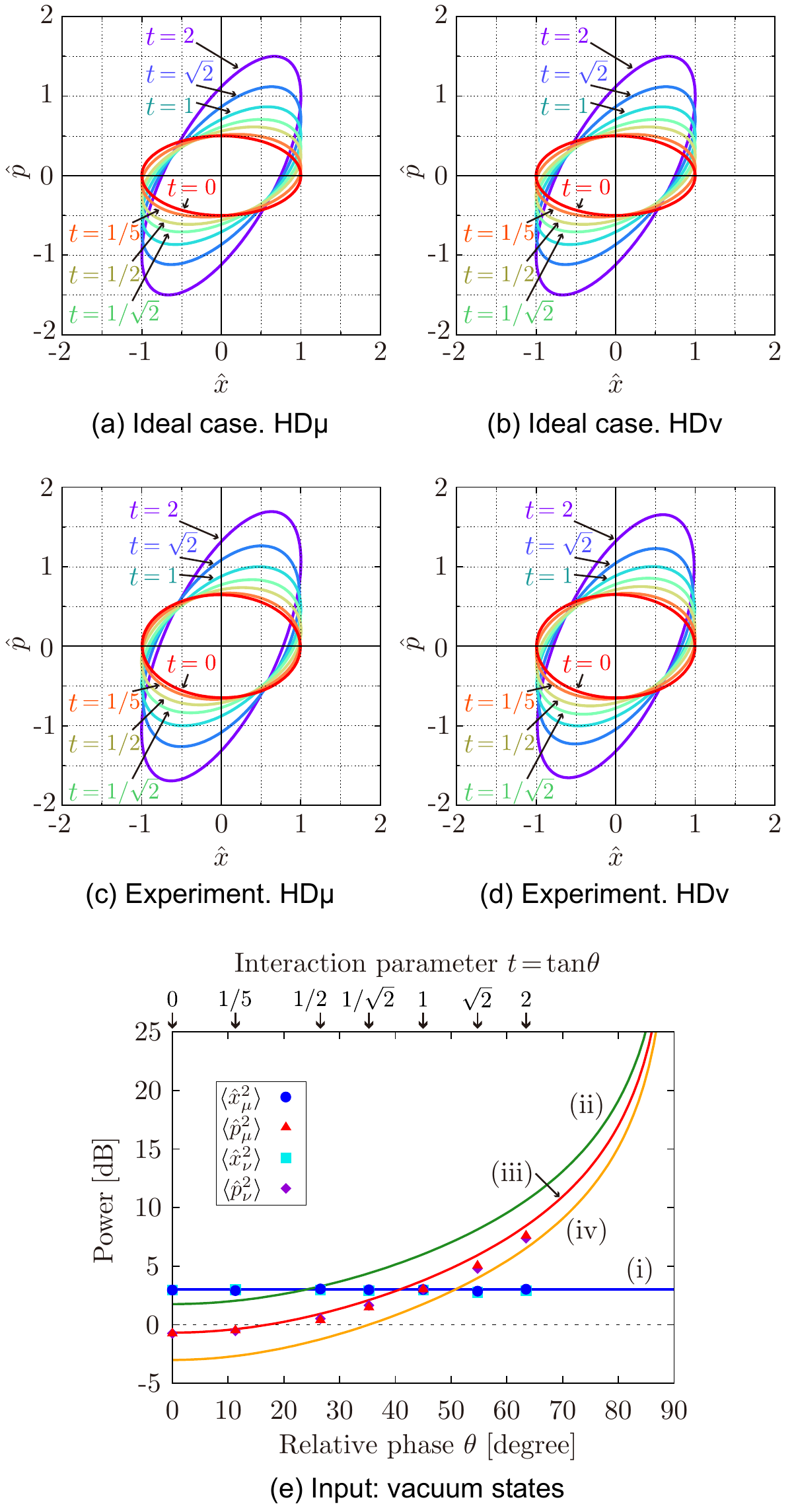}
			\caption{
(Color online) Output states for a \Tz gate operating with several interaction parameters $t=\tan\theta\in\{0,\, 1/5,\, 1/2,\, 1/\sqrt{2},\, 1,\, \sqrt{2},\, 2\}$, employing two vacuum inputs.
(a)--(d) Phase-space distributions. The second moments of Gaussian Wigner functions are represented by ellipses.
(a) and (b) Theoretical predictions for the ideal case with infinite resource squeezing.
(c) and (d) Experimental results computed from the measured variances of $\hat x_j$, $\hat p_j$, and $(\hat x_j\pm\hat p_j)/\sqrt{2}$, where $j\in\{\mu, \nu\}$.
(e) The measured variances of $\hat x_j$, $\hat p_j$.
The horizontal axis is the relative phase $\theta$ between the signal beam and the local oscillator beam at HD-2, which determines the  interaction parameter, as $t=\tan \theta$.
The coloured lines show the theoretical predictions of $\langle\hat x^2_j\rangle$ (i), $\langle\hat p^2_j\rangle$ without squeezing (ii), $\langle\hat p^2_j\rangle$ with $-4.5$~dB resource squeezing (iii), and $\langle\hat p^2_j\rangle$ with inifinite squeezing (unphysical, ideal case) (iv).
Error bars are omitted, because they are very small compared to the scale of the vertical axis.
}
			\label{fig:PhaseSpace}
\end{figure}

In \refFigs{PhaseSpace}{(a)}{(d)}, we visualise the phase-space distributions of the output Gaussian states by ellipses
for seven different interaction parameters $t=\tan\theta\in\{0,\, \tfrac{1}{5},\, \tfrac{1}{2},\, \tfrac{1}{\sqrt{2}},\, 1,\, \sqrt{2},\, 2\}$, 
for vacuum state inputs. These interaction parameters correspond to the following measurement angles, $\theta\in\{0.0^\circ,\, 11.3^\circ,\, 26.6^\circ,\, 35.3^\circ,\, 45.0^\circ,\, 54.7^\circ,\, 63.4^\circ\}$.
The second moments are expressed by the size of the phase-space ellipse, which corresponds to the cross section of the quantum state's Wigner function.
Local and short radii correspond to $\sqrt{2}$ times standard deviations in the corresponding directions.

The theoretical predictions of the ideal case with infinite resource squeezing ($r\to\infty$) are shown in \refFigsAnd{PhaseSpace}{(a)}{(b)}.
Here, we see that the $\hat{x}$ quadrature amplitudes remain fixed, while the $\hat{p}$ quadrature amplitudes increase with larger interaction parameter values.
The broadening in $\hat{p}$ is due to the uncorrelated quantum fluctuations of both $\hat{x}_\alpha$ and $\hat{x}_\beta$ being added to $\hat{p}_\alpha$ and $\hat{p}_\beta$ by the interaction parameter dependent $T_\mathrm{Z}$ gate. Note that the additional local squeezing operations decrease these fluctuations.
The variances of $\hat x_\alpha$ and $\hat x_\beta$ are fixed at twice the shot noise level (SNL) from the additional local squeezing and are not dependent on the interaction parameter.

The experimental results are shown in \refFigsAnd{PhaseSpace}{(c)}{(d)}, which are
calculated from the measured variances of $\hat x_j$, $\hat p_j$, and $(\hat x_j\pm\hat
p_j)/\sqrt{2}$, where $j\in\{\mu, \nu\}$. We assume a Gaussian distribution and zero mean
value. Each of the two output modes have a nearly identical phase-space distribution with respect to
each other, indicative of the high level of symmetry in our optical mode matching. We see a slight
broadening in $\hat p$ compared to the ideal case predicted by theory, due to the finite resource
squeezing which couples in excess noise, while $\hat{x}$ remains unaffected, in accordance with $\hat{\bm{\delta}}$ in Eq.\eqref{delta}.

In order to compare them with the following results, the measured variances of $\hat x_\alpha$, $\hat x_\beta$, $\hat p_\alpha$, and $\hat p_\beta$ are plotted in \refFig{PhaseSpace}(e).
The horizontal axis is the relative phase $\theta$ between the signal beam and the local oscillator beam at HD-2.
The variances of $\hat x$ are 3.0~dB above the SNL independent of the resource squeezing level $r$ and the interaction parameter $\tan\theta$ as expected from the theory expressed by the blue line (i),while $\hat p$ depends on them.
The green line (ii) represents the theoretical predictions for zero resource squeezing, while the orange line (iv) represents infinite squeezing. Finite squeezing values appear between these two extremes, and we find our experimental results are close to the theoretical prediction of $-4.5$~dB resource squeezing, as indicated by the red line (iii). These results indicate a good qualitative agreement with the theoretical predictions.

\begin{figure}
		\centering
		\includegraphics[scale=0.65,clip]{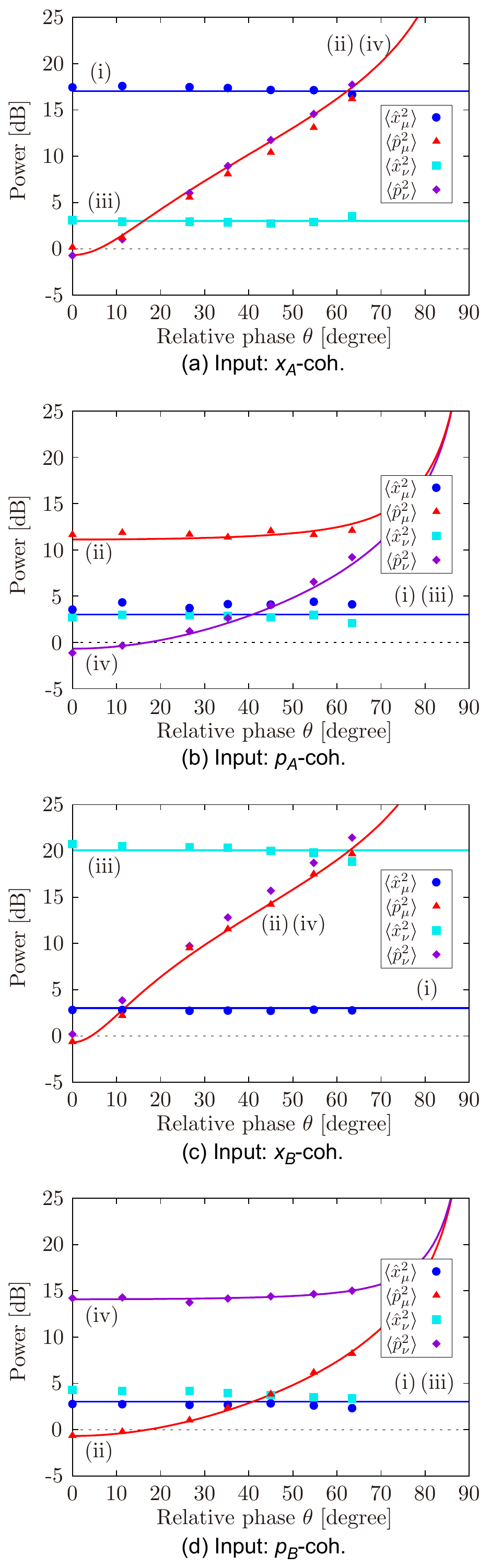}
			\caption{
(Color online) The powers at the outputs from two coherent inputs with several interaction parameters.
The horizontal axes are the relative phases $\theta$ between the signal beam and the local oscillator beam at HD-2, which are related to the interaction parameters $\tan \theta$.
(a)--(d) $(\langle\hat x_\alpha\rangle,\langle\hat p_\alpha\rangle,\langle\hat x_\beta\rangle,\langle\hat p_\beta\rangle)$ are $(a,0,0,0)$, $(0,a,0,0)$, $(0,0,b,0)$, and $(0,0,0,b)$,
where $a^2$ and $b^2$ correspond to 13.8~dB and 16.9~dB above the shot noise level, respectively.
(i), (ii), (iii), and (iv) show the theoretical predictions of $\langle\hat x^2_\mu\rangle$, $\langle\hat p^2_\mu\rangle$, $\langle\hat x^2_\nu\rangle$, and $\langle\hat p^2_\nu\rangle$ with $-4.5$~dB resource squeezing, respectively;
coh., coherent state.
Error bars are omitted, because they are very small compared to the scale of vertical axis.
}
			\label{fig:coherent}
\end{figure}

Next, we replace one of the input vacuum states by coherent states, allowing us to verify the input-output relationship based on the assumption that the gate has a linear response.
The powers of the input amplitude quadratures are individually measured in advance, corresponding to 13.8~dB for mode $\alpha$ and 16.9~dB for mode $\beta$, respectively, compared to the SNL.

In analogy with \refFig{PhaseSpace}, \refFigure{coherent}(a) shows the powers of the output quadratures for an input coherent state $\alpha$ and an input vacuum state $\beta$.
The output quadrature powers are shown as a function of the relative phase $\theta$ between the signal beam and the local oscillator beam at HD-2, which determines the interaction parameter $\tan\theta$.  Theoretical predictions are shown as lines and experimental data as markers. The predictions are calculated from the measured input coherent amplitude with a resource squeezing level of $-4.5$~dB.
We observe fixed power increases in $\hat{x}$ and $\theta$-dependent increases in $\hat{p}$. The power of $\hat{x}_\mu$ increases by 3.0~dB above the inital 13.8~dB (corresponding to about 17~dB above the SNL, blue markers), which is due to the additional local squeezing operation. The power of $\xx_\nu$ is the same as the case of two vacuum inputs (corresponding to 3.0~dB above the SNL, cyan markers).
$\pp_\mu$ and $\pp_\nu$ experience larger increases in power relative to the case of vacuum inputs in \refFig{PhaseSpace}(e), due to the increasing contribution of the nonzero coherent amplitude of $\xx_\alpha$ via the \Tz gate.
Similarly, \refFigures{coherent}{(b)}{(d)} show the results for a nonzero coherent amplitude in the $\hat{p}_\alpha$, $\hat{x}_\beta$, and $\hat{p}_\beta$ input quadratures, respectively. The \Tz gate behaves as predicted, with the sum of $\hat{x}_\alpha$ and $\hat{x}_\beta$ appropriately appearing in both $\hat{p}_\mu$ and $\hat{p}_\nu$ quadratures, as a function of the interaction parameters.
The small discrepancies between our experimental results and the theoretical predictions are caused by the (slightly unbalanced) propagation losses and non-unity homodyne detections.

\begin{figure}[t!]
\centering
\includegraphics[scale=0.85,clip]{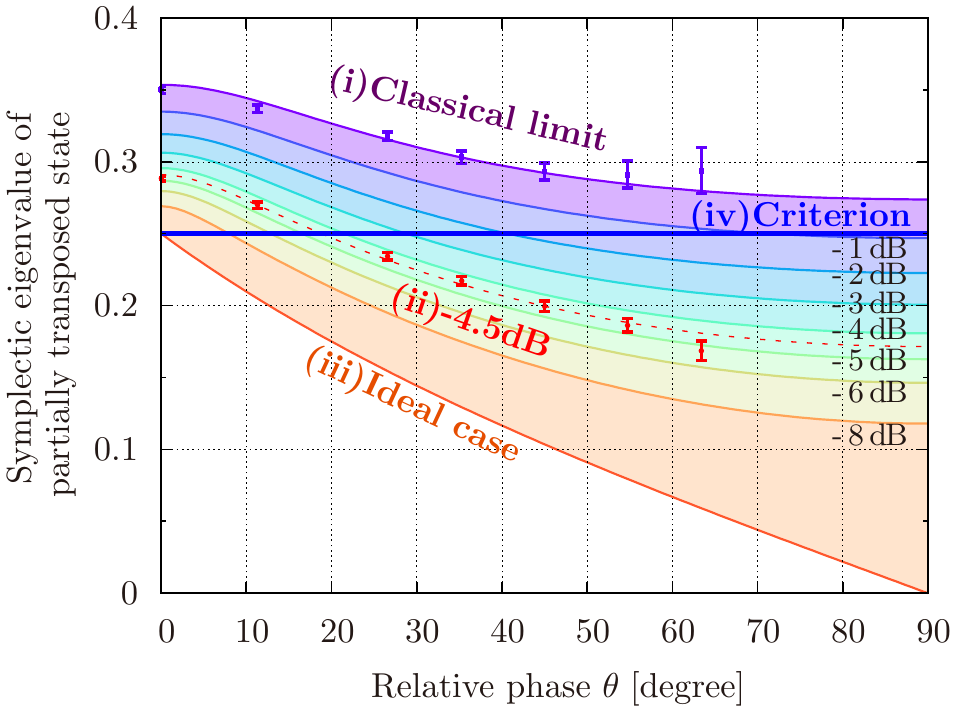}
\caption{
(Color online)
The dependence of the entanglement at the outputs on the interaction parameters of the \Tz gate.
The horizontal axis corresponds to the relative phase $\theta$ between the signal beam and the local oscillator beam at HD-2, which determines the interaction parameter $\tan \theta$.
The vertical axis corresponds to the symplectic eigenvalues of the partially transposed covariance matrix of the output state, connected to a measure of entanglement (see text).
(i) without squeezing, (ii) with $-4.5$~dB resource squeezing, (iii) with infinite squeezing (unphysical, ideal case) and (iv) quantum boundary; values below satisfy a sufficient condition for entanglement.
Error bars show standard errors.
}
\label{fig:entanglement}
\end{figure}

Finally, the entanglement strength is quantified in \refFig{entanglement}. Shown there is the set of symplectic eigenvalues $\tilde{\lambda}_-$ of the partially transposed covariance matrices corresponding to the output states. These are calculated from the variances of the output quadratures for vacuum inputs (see Ref.~\cite{Yokoyama14} for details), and are displayed as a function of \Tz interaction parameter (as determined by the relative phase of homodyne detection).
Note that the results of covariance matrices satisfy the physicality condition $V+(i/4)\ \Omega \ge 0$, where
$\Omega $ is a direct sum of
$\left(\begin{smallmatrix}
0 & -1 \\
1 & 0
\end{smallmatrix}
\right)$ \cite{Simon94,Pirandola09}.
The theoretical predictions for the experiment with and without resource squeezing are represented by the theoretical curves (ii) and (i), respectively.
We observe the remarkable feature of an enhancement in entanglement strength dependent on the interaction parameter. The entangling criterion is satisfied for parameter values of $\tan\theta=\tfrac{1}{2}$, $\frac{1}{\sqrt{2}}$, 1, $\sqrt{2}$, and 2 when the resource state is squeezed. Conversely, without squeezing the symplectic eigenvalues never cross the quantum boundary for any value of interaction parameter.

\section{conclusion}
In conclusion, we have proposed and experimentally demonstrated a fully tuneable \Tz gate for continuous-variable one-way quantum computation. Our proof-of-principle demonstration employed a three-mode linear cluster state as a resource for implementing a new cluster gain tuning protocol.
The capability of the gate to produce entanglement at the output is verified via the symplectic eigenvalues of the partially transposed covariance matrix of the output for the case of two coherent input states.
The interaction parameter at the gate and accordingly the entanglement strength in the output state are accurately tuned by a corresponding tuning of the set of measurement bases. Since our gate can be directly incorporated into large-scale one-way quantum computation schemes, it may facilitate efficient implementations of MBQC with cluster states.

\section*{ACKNOWLEDGMENTS}
This work was partly supported by PDIS, GIA, APSA, and FIRST initiated by CSTP,
ASCR-JSPS, and SCOPE program of the MIC of Japan.
S.Y.\ acknowledges support from ALPS\null.
R.U.\ acknowledges support from JSPS\null.
S.A.\ acknowledges support from the Prime Minister's Award.

\appendix
\section{Mathematical treatment of feedforward}
\begin{figure}[b]
		\centering
		\includegraphics[scale=1.4,clip]{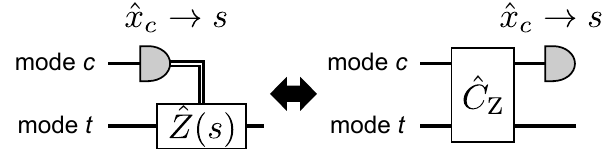}
			\caption{
Equivalent quantum circuits.
(left) Executing displacement after measurement.
(right) Measurement after an interaction gate.
}
			\label{fig:appendix}
\end{figure}
We consider a situation where some observable $\hat{s}_c$ of a control mode $c$ is measured and then the outcome $s$ is used for feedforward to a target mode $t$ as a displacement operation $\hat{Z}_t(s)=e^{2is\hat{x}_t}$. 
As its name suggests, a displacement operator $\hat{Z}_t(s)$ displaces a quadrature operator in $\hat{p}$ direction, 
\begin{align}
\hat{Z}_t^\dagger(s)\hat{x}_t\hat{Z}_t(s)=&\hat{x}_t, &
\hat{Z}_t^\dagger(s)\hat{p}_t\hat{Z}_t(s)=&\hat{p}_t+s.
\end{align}
It is a well-known fact, as depicted in \refFig{appendix}, that a measurement in the middle of successive unitary gates can be moved to the last by appropriately replacing the feedforward circuits by controlled gates.
Based on this equivalence, here as a matter of notation, we write a $\hat{Z}_t(s)$ gate dependent on a measurement outcome $s$ of an observable $\hat{s}_c$ as $\hat{Z}_t(\hat{s}_c)=e^{2i\hat{s}_c\hat{x}_t}$, which transforms the quadrature operators, 
\begin{align}
\hat{Z}_t^\dagger(\hat{s}_c)\hat{x}_t\hat{Z}_t(\hat{s}_c)=&\hat{x}_t, &
\hat{Z}_t^\dagger(\hat{s}_c)\hat{p}_t\hat{Z}_t(\hat{s}_c)=&\hat{p}_t+\hat{s}_c.
\end{align}
As a special case, when $\hat{s}_c = \hat{x}_c$, the equivalent gate $\hat{Z}_t(\hat{x}_c)=e^{2i\hat{x}_c\hat{x}_t}=\hat{C}_{Zct}$ is a controlled phase gate.

Similarly, we can also consider the case of $\hat{x}$-direction displacement feedforward $\hat{X}_t(s)=e^{-2is\hat{p}_t}$, where the equivalent controlled gate is $\hat{X}_t(\hat{s}_c)=e^{-2i\hat{s}_c\hat{p}_t}$.

\section{Nullifiers after cluster gain tuning}
In the cluster gain tuning starting from a three-mode linear cluster state in \refFig{ExperimentalSetup}(a), first a center mode is measured with respect to an observable, 
\begin{align}
\hat{s}_2(\theta) = \hat{x}_{C2}\cos\theta-\hat{p}_{C2}\sin\theta.
\end{align}
Then the outcome $s_2$ is used for a feedforward displacement operation $\hat{Z}_{C1}(-s_2/\cos\theta)\hat{Z}_{C3}(-s_2/\cos\theta)$, which transforms the quadratures of the remaining two modes as,
\begin{subequations}
\begin{align}
\hat{x}_{C1}^\prime 
& = \hat{Z}_{C1}^\dagger\left(-\frac{\hat{s}_2(\theta)}{\cos\theta}\right)\hat{x}_{C1}\hat{Z}_{C1}\left(-\frac{\hat{s}_2(\theta)}{\cos\theta}\right) \notag\\
& = \hat{x}_{C1} \\
\hat{p}_{C1}^\prime 
& = \hat{Z}_{C1}^\dagger\left(-\frac{\hat{s}_2(\theta)}{\cos\theta}\right)\hat{p}_{C1}\hat{Z}_{C1}\left(-\frac{\hat{s}_2(\theta)}{\cos\theta}\right) \notag\\
& = \hat{p}_{C1} - \frac{\hat{s}_2(\theta)}{\cos\theta} \notag\\
& = \hat{p}_{C1} - \hat{x}_{C2} + \hat{p}_{C2}\tan\theta \\
\hat{x}_{C3}^\prime 
& = \hat{Z}_{C3}^\dagger\left(-\frac{\hat{s}_2(\theta)}{\cos\theta}\right)\hat{x}_{C3}\hat{Z}_{C3}\left(-\frac{\hat{s}_2(\theta)}{\cos\theta}\right) \notag\\
& = \hat{x}_{C3} \\
\hat{p}_{C3}^\prime 
& = \hat{Z}_{C3}^\dagger\left(-\frac{\hat{s}_2(\theta)}{\cos\theta}\right)\hat{p}_{C3}\hat{Z}_{C3}\left(-\frac{\hat{s}_2(\theta)}{\cos\theta}\right) \notag\\
& = \hat{p}_{C3} - \frac{\hat{s}_2(\theta)}{\cos\theta} \notag\\
& = \hat{p}_{C3} - \hat{x}_{C2} + \hat{p}_{C2}\tan\theta. 
\end{align}
\end{subequations}
Therefore, bearing in mind the nullifiers for the initial quadratures, 
\begin{subequations}
\begin{align}
\hat{\delta}_1 & = \hat{p}_{C1}-\hat{x}_{C2}, \\
\hat{\delta}_2 & = \hat{p}_{C2}-\hat{x}_{C1}-\hat{x}_{C3}, \\
\hat{\delta}_3 & = \hat{p}_{C3}-\hat{x}_{C2}, 
\end{align}
\end{subequations}
the nullifiers for the new quadratures are constructed only from the quadrature operators of the modes $C1$ and $C3$ as
\begin{subequations}
\begin{align}
\hat{\delta}_1^\prime 
& = \hat{p}_{C1}^\prime-(\hat{x}_{C1}^\prime+\hat{x}_{C3}^\prime)\tan\theta \notag\\
& = (\hat{p}_{C1}-\hat{x}_{C2}) + (\hat{p}_{C2}-\hat{x}_{C1}-\hat{x}_{C3})\tan\theta \notag\\
& = \hat{\delta}_1 + \hat{\delta}_2\tan\theta, \\
\hat{\delta}_3^\prime 
& = \hat{p}_{C3}^\prime-(\hat{x}_{C1}^\prime+\hat{x}_{C3}^\prime)\tan\theta \notag\\
& = (\hat{p}_{C3}-\hat{x}_{C2}) + (\hat{p}_{C2}-\hat{x}_{C1}-\hat{x}_{C3})\tan\theta \notag\\
& = \hat{\delta}_3 + \hat{\delta}_2\tan\theta. 
\end{align}
\end{subequations}

\end{document}